\documentclass[preprint,12pt]{elsarticle}

\usepackage{amssymb}

\usepackage{longtable}
\usepackage{float}
\usepackage{subfigure}
\usepackage{MnSymbol}
\usepackage{graphicx}
\usepackage{adjustbox}
\usepackage{array}
\usepackage{multirow}
\usepackage{caption}
\usepackage{tabu}
\usepackage{tabularx}
\usepackage{makecell}
\usepackage{color,soul}
\usepackage{booktabs}

\usepackage[hyphens]{url}

\begin{document}

\begin{frontmatter}

\title{Assessing the Reproducibility of Machine-learning-based Biomarker Discovery in Parkinson's Disease}

\author[CS]{Ali Ameli}

\author[CS,Bio]{Lourdes Peña-Castillo}
\ead{lourdes@mun.ca}

\author[CS,Math]{Hamid Usefi}
\ead{usefi@mun.ca}

\affiliation[CS]{organization={Department of Computer Science, Memorial University of Newfoundland},
            addressline={230 Elizabeth Ave}, 
            city={St. John's},
            postcode={A1C5S7}, 
            state={NL},
            country={Canada}}
            
\affiliation[Bio]{organization={Department of Biology, Memorial University of Newfoundland},
              addressline={230 Elizabeth Ave}, 
            city={St. John's},
            postcode={A1C5S7}, 
            state={NL},
            country={Canada}}

\affiliation[Math]{organization={Department of Mathematics and Statistics, Memorial University of Newfoundland},
              addressline={230 Elizabeth Ave}, 
            city={St. John's},
            postcode={A1C5S7}, 
            state={NL},
            country={Canada}}

\begin{abstract}

Genome-Wide Association Studies (GWAS) help identify genetic variations in people with diseases such as Parkinson's disease (PD), which are less common in those without the disease.  Thus, GWAS data can be used to identify genetic variations associated with the disease.  Feature selection and machine learning approaches can be used to analyze GWAS data and identify potential disease biomarkers. However, GWAS studies have technical variations that affect the reproducibility of identified biomarkers, such as differences in genotyping platforms and selection criteria for individuals to be genotyped. To address this issue, we collected five GWAS datasets from the database of Genotypes and Phenotypes (dbGaP) and explored several data integration strategies. We evaluated the agreement among different strategies in terms of the Single Nucleotide Polymorphisms (SNPs) that were identified as potential PD biomarkers. Our results showed a low concordance of biomarkers discovered using different datasets or integration strategies. However, we identified fifty SNPs that were identified at least twice, which could potentially serve as novel PD biomarkers. These SNPs are indirectly linked to PD in the literature but have not been directly associated with PD before. These findings open up new potential avenues of investigation.
\end{abstract}

\begin{highlights}
\item Assessed the reproducibility of Parkinson's biomarkers discovery using  feature selection,  machine learning and four dataset integration strategies. 
\item There is low concordance of biomarkers discovered using different datasets or integration strategies.
\item On average 93\% of the SNPs discovered using a single data set are not replicated in other datasets.
\item Dataset integration reduces  the lack of replication of SNPs discovered by a 66\% (from 93\% to 62\%).
\item We found 50 replicated SNPs indirectly associated with PD. That is, they are directly associated in the literature with a disease that frequently co-occurs with PD. 
\end{highlights}


\end{frontmatter}

\section{Introduction}\label{sec1}

Parkinson's disease (PD) is a degenerative neurological condition that affects both the motor and non-motor aspects of movement including planning, initiation, and execution \citep{jankovic2008parkinson,  contreras1995effects}.   Parkinson's symptoms result from an 80\% or greater loss of dopamine-producing cells in the substantia nigra~\citep{Dopamine}. Dopamine works with other neurotransmitters to coordinate nerve and muscle cells involved in movement~\citep{Dopamine}. Without sufficient dopamine, the neurotransmitters' balance is disrupted, causing the distinctive symptoms of PD, such as tremors, rigidity, slow movement, and poor coordination~\citep{opara2012quality}.  PD patients experience a significant decline in quality of life, social activities, and family relationships~\citep{johnson2013economic, kowal2013current, yang2017economic}.  PD is one of the most prevalent neurodegenerative disorders affecting 1-2 persons per 1,000 people at any time of their live and has a prevalence rate of 1\% over the age of 60 \citep{tysnes2017epidemiology}. Due to a growth in the number of senior individuals and age-standardized incidence rates, the estimated number of people affected with PD in the world grew from 2.5 million in 1990 to 6.1 million in 2016~\citep{dorsey2018global}.  

Motor symptoms are mostly used to diagnose PD~\citep{Dopamine}.  While non-motor symptoms (such as cognitive changes, difficulties with attention and planning, sleep disorders, sensory abnormalities, and olfactory dysfunction) are common in patients before the onset of PD, they lack specificity, are challenging to assess, and/or vary from patient to patient \citep{jankovic2008parkinson, tremblay2017trigeminal, zesiewicz2006nonmotor}. Thus non-motor symptoms are not currently utilized to diagnose PD on their own, despite some of them being used as supportive diagnostic criteria \citep{braak2003staging, postuma2015mds}.  Additionally, there is no specific lab test to diagnose PD~\citep{Dopamine}.  

Machine Learning (ML) techniques have already been applied to a variety of data to generate models to support PD diagnosis.  Types of data used to generate ML models for PD diagnosis include handwriting patterns \citep{drotar2014decision, pereira2018handwritten}, gait~\citep{li2022detecting}, movement \citep{wahid2015classification, pham2017tensor}, neuroimaging \citep{cherubini2014magnetic, choi2017refining,  segovia2019assisted}, speech \citep{sakar2013collection, ma2014efficient}, Cerebrospinal Fluid (CSF) \citep{lewitt20133, maass2020elemental}, cardiac scintigraphy \citep{nuvoli2020123i}, serum \citep{varadi2019serum}, gene expression patterns from microarrays \citep{su2020mining}, and Optical Coherence Tomography (OCT) \citep{nunes2019retinal}.  ML has also been used in  the combination of several data types, such as Magnetic Resonance Imaging (MRI) and Single-Photon Emission Computed Tomography (SPECT) data \citep{cherubini2014magnetic, wang2017multi}. 

Genome-Wide Association Studies (GWAS) aim to identify genetic variations, especially Single Nucleotide Polymorphisms (SNPs), that distinguish individuals with a particular disease from those without it. By analyzing these differences,  one can identify SNPs that are associated with the disease of interest.  However, GWAS data are massive and contain more SNPs (features) than individuals (samples), which poses a challenge of high dimensionality. To address this, feature selection is a standard technique used to choose the most reliable, non-redundant, and relevant features to include in a model. The main goals of feature selection are to enhance the predictive model's performance and reduce computational expenses. In this study,  we utilized five different GWAS datasets that included both Parkinson's disease (PD) cases and controls. To reduce the dimensionality of these datasets, we employed the SVFS algorithm \citep{afshar2021dimensionality} for feature selection. Additionally, we used the Random Forest (RF) algorithm~\citep{breiman2001random} for classification and proposed four different approaches to integrate datasets, which we compared with the baseline approach of no integration.  Finally,  we collected SNPs that were identified by at least two different datasets or integration strategies.   We identified  four SNPs with direct links to PD and 50 with indirect links to PD in the literature.  A direct link means that current literature directly links a SNP with PD; while an indirect link means that current literature suggests the involvement of a SNP in a disease other than PD but this other disease co-occurs with PD in a significant number of PD patients.  These indirectly-associated SNPs open up new potential avenues for investigating potential biomarkers for PD.

\section{Methods}\label{sec3}

\subsection{Dataset Description}\label{subsec1}
Our study includes five different GWAS datasets obtained from dbGaP~\citep{dbgap}. A description of these datasets is as follows and a short summary is given in Table \ref{table:1}. 
\begin{enumerate}
    \item Phs000126 (Familial) \citep{Familial, nichols2007lrrk2, pankratz2006mutations, nichols2005genetic, wilk2006herbicide, sun2006influence, karamohamed2005bdnf, karamohamed2005absence} dataset combines the results of two major National Institutes of Health (NIH)-funded genetic research studies (PROGENI and GenePD) which aimed at discovering new genes that influence the risk of PD.  These studies have been analyzing and recruiting families with two or more PD affected members for over eight years. There are almost 1,000 PD families in the total sample.
       
 \item Phs000394 (Autopsy)-Confirmed Parkinson Disease GWAS Consortium (APDGC) \citep{Autopsy} was established to perform a genome-wide association research in people with neuropathologically diagnosed PD and healthy controls. Their study's hypothesis is that by enrolling only cases and controls with neuropathologically proven illness status, diagnostic misclassification will be reduced and power to identify novel genetic connections will be increased.

\item Phs000089 The National Institute of Neurological Disorders and Stroke (NINDS) repository \citep{NINDS, fung2006genome, simon2007genome, simon2009genome, sanz2020association}  was created in 2001 with the intention of creating standardized, widely applicable diagnostic and other clinical data as well as a collection of DNA and cell line samples to enhance the field of neurological illness gene discovery.  NINDS dataset is divided into NINDS1 and NINDS2.

\item Phs000048 (Tier 1)  The dbGaP team at NCBI calculated this Genome-Wide Association scan \citep{Tier1, maraganore2005high, evangelou2007meta, lesnick2007genomic} between genotype and PD status of 443 sibling pairs that were at odds for PD between June 1996 and May 2004.

    \end{enumerate}

\begin{table}[h]
\centering
\caption{Dataset description}\label{table:1}
\resizebox{\textwidth}{!}{%
    \begin{tabular}{@{}lccccc@{}} 
     \toprule
     Dataset ID & \#  Samples & \#  Missing Phenotype & \#  Cases & \# Controls & \# SNPs \\
     \midrule
     1. phs000394 (Autopsy) & 1001 & 24 & 642 & 335 & 1134514 \\ 
     2. phs000126 (Familial) & 2082 & 315 & 900 & 867 & 344301\\
     3. phs000089 (NINDS1) & 1741 & 0 & 940 & 801 & 545066\\ 
     4. phs000089 (NINDS2) & 526 & 0 & 263 & 263 & 241847\\
     5. phs000048 (Tier1) & 886 & 0 & 443 & 443 & 198345\\
      \bottomrule
    \end{tabular}
    }
\end{table}

\subsection{Preprocessing and imputation}\label{subsubsec1}

We applied KNNcatimputer \citep{schwender2012imputing} for imputing missing values. 
To tune the parameters of 
KNNcatimputer using grid search, we removed from 5\% to 20\%  of genotype values from the data, impute them and evaluate the imputation accuracy. We found that  ``Cohen” for  distance measure and $n = 20$ for the number of neighbours as the parameters to optimize imputation accuracy. We also removed columns (SNPs) with more than $5\%$ missing values. We came up with the threshold of  $5\%$ by experimenting on columns without missing values, randomly removed a certain percentage of the values, and used KNNcatimputer to determine the accuracy of imputation. On the Familial, Autopsy, NINDS1, NINDS2, and Tier1 datasets, the accuracy of imputation for different parameters was comparable to one another. In order to get consistent findings across all datasets, we chose the aforementioned parameters. The imputation accuracy for Familial, Autopsy, NINDS1, NINDS2, and Tier1 dataset was $84.04\%\pm1.12$, $87.38\%\pm2.31$,  $86.31\%\pm1.5$, $84.53\%\pm1.3$, and $83.81\%\pm0.42$ respectively.

\subsection{Feature Selection}\label{subsec4}

We used the SVFS algorithm \citep{afshar2021dimensionality} for dimensionality reduction. Afshar and Usefi demonstrated the efficiency of SVFS  in terms of accuracy and running time on various biological datasets, and compared it against other feature selection algorithms. As per  \citep{afshar2021dimensionality} we set SVFS parameters as 
$k= 50$, $Th_{irr}= 3$, $Th_{red}= 4$, $\alpha= 50$, and $\beta= 5$ and we use these values on all datasets and experiments.

\subsection{Classification}

For classification,  we used the Random Forest (RF) algorithm~\citep{breiman2001random}. We performed repeated 5-fold cross-validation (CV) for 10 times. We used RF with the following settings: n\_estimator = 100, criteria = “gini”, max depth = None, and min\_samples\_split = 2.

\subsection{Common SNPs among datasets}\label{subsec4.5}
We observed that the Familial dataset  has the most number of samples  and SNPs (Table~\ref{table:1}).  As we discuss in Section~\ref{subsec5},  for pair-wise dataset integration, we need to obtain the SNPs in common between two datasets. Thus, to maximize the number of SNPs in common between two datasets we decided to use the Familial dataset as the dataset to be integrated with all other datasets. The number of SNPs in common between the Familial and all other datasets are provided in Table~\ref{commonSNPs}.
\begin{table}[h]
\centering
\caption{Number of SNPs in common between Familial and the other four  datasets}\label{commonSNPs}%
\resizebox{\columnwidth}{!}{%
\begin{tabular}{@{}llc@{}}
\toprule
Dataset  1 & Dataset  2  & \# SNPs in common\\
\midrule
     phs000126 (Familial) & phs000394 (Autopsy) & 209,730\\ 
     phs000126 (Familial) & phs000089 (NINDS1) & 305,812\\
     phs000126 (Familial) & phs000089(NINDS2) & 4,034\\
     phs000126 (Familial) & phs000048(Tier1) & 28,646\\
 \bottomrule
\end{tabular}%
}
\end{table}

\subsection{Approaches to integrate datasets}\label{subsec5}
We proposed four different approaches to integrate datasets, and compared these approaches against the baseline approach of  no integration (Approach 0). For each approach, we defined two modes: (A) take the most-frequently-selected SNPs as features for the Random Forest algorithm, and (B) extend the SNPs in A by adding those SNPs in linkage disequilibrium (LD) with the SNPs in A and used all these SNPs as features for the Random Forest algorithm (RF).  We used SNIPA \citep{SNIPA} to get SNPs in LD with the SNPs in A.  SNIPA parameters:  genome assembly, variant set, population, and genome annotation were set as  GRCH-37, 1000 genomes phase 3 V5, European, and Ensemble 87, respectively, throughout all our experiments.  The most-frequently-selected SNPs were defined as those SNPs selected in at least fives runs out of 50 runs of the SVFS algorithm. These SNPs are referred to as  the \textit{most common SNPs}.

\subsubsection{Approach 0}\label{subsubsec3}

Approach 0 is our baseline approach. In this approach we ran the SVFS algorithm 50 rounds (5-fold CV for 10 times) on each dataset separately to find the most common SNPs per dataset.  We used cross-validation to assess the classification performance of  a model generated using the most  common SNPs as features and RF as the classifier. This process was performed again following mode B described above.

\subsubsection{Approach 1}\label{subsubsec4}
In Approach 1, we selected features from the Familial dataset and found the SNPs to use for classification as per mode A and B above. Then, for each of the other four datasets, found which of the selected SNPs was available on that specific dataset and performed CV to train and assess the classification performance of a model generated using that dataset and those SNPs.

\subsubsection{Approach 2}\label{subsubsec5}
In Approach 2, we first obtained the intersection of SNPs between the Familial dataset and each of the other four datasets (Table~\ref{commonSNPs}). SNPs not in the intersection were removed from both datasets. We did then followed the  same steps as per Approach 1 by selecting the features from the condensed version of the Familial dataset and doing CV on each of the other datasets. 

\subsubsection{Approach 3}\label{subsubsec6}
In Approach 3, before doing feature selection, we got the SNPs in the intersection between the Familial dataset and the other 4 datasets, and then merged the datasets  pair-wise as follows: Familial and Autopsy, Familial and NINDS1, Familial and NINDS2, and Familial and Tier1. We ran the SVFS feature selection algorithm on each of the four merged datasets and extracted the most common SNPs. Then, we performed CV to assess the classification performance of a model generated using each of the four merged datasets.

To allow for a direct comparison with the other approaches, in this approach we calculated the accuracy per each dataset in addition to the accuracy on the merged dataset. Thus, we obtained three accuracies: one for the merged dataset and one for each of the merged dataset's individual instances. 

\subsubsection{Approach 4}\label{subsubsec7}
This approach is the same as Approach 3, but equal number of samples  from each of the two datasets were randomly selected before merging them. The number of cases and controls taken from each dataset is the same.

\section{Results and Discussion}\label{sec4}
\begin{figure}[h]
    \centering
    \subfigure[Approach 0 (A)]{\includegraphics[width=0.45\textwidth]{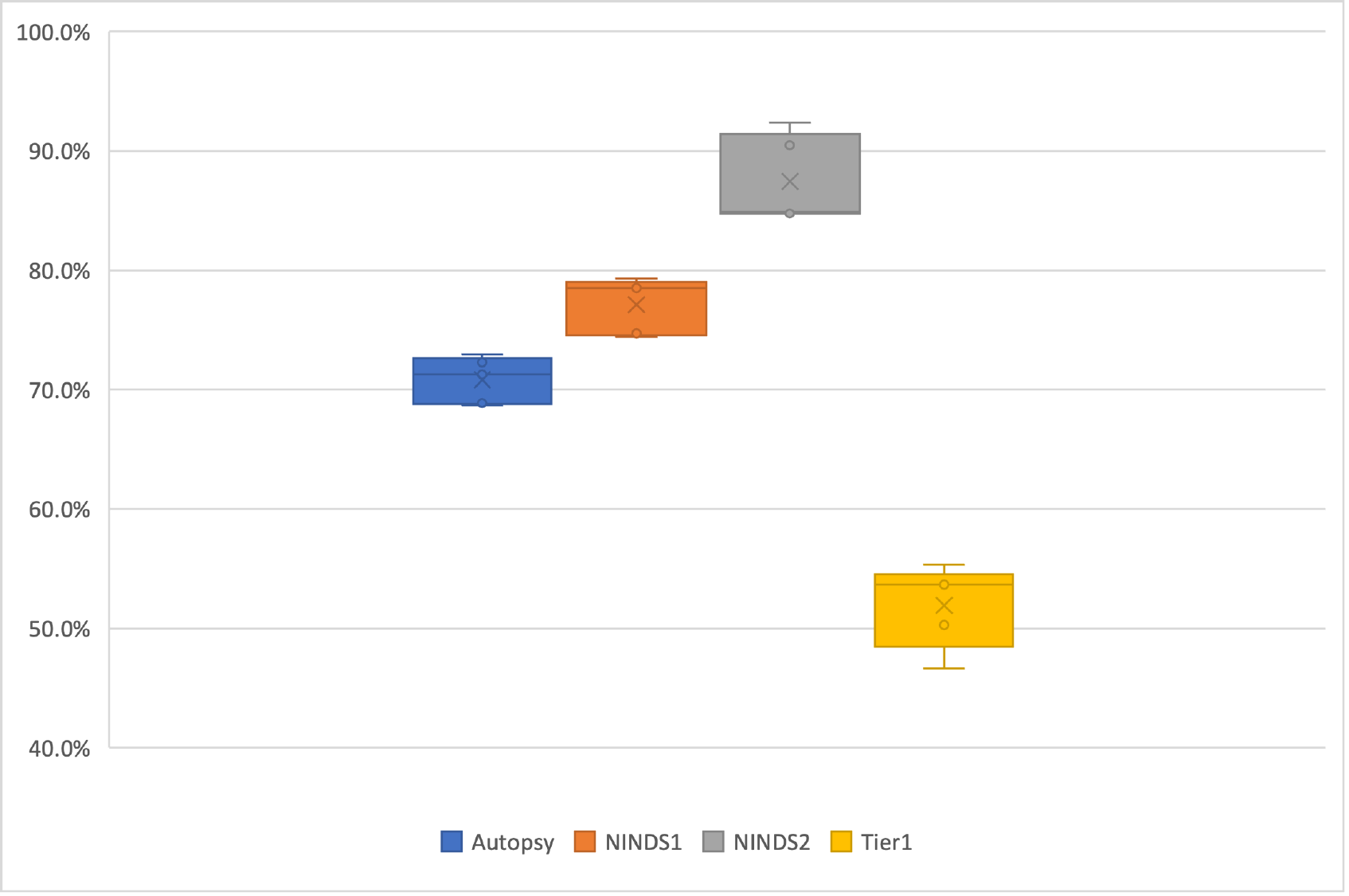}}
    \subfigure[Approach 0 (B)]{\includegraphics[width=0.45\textwidth]{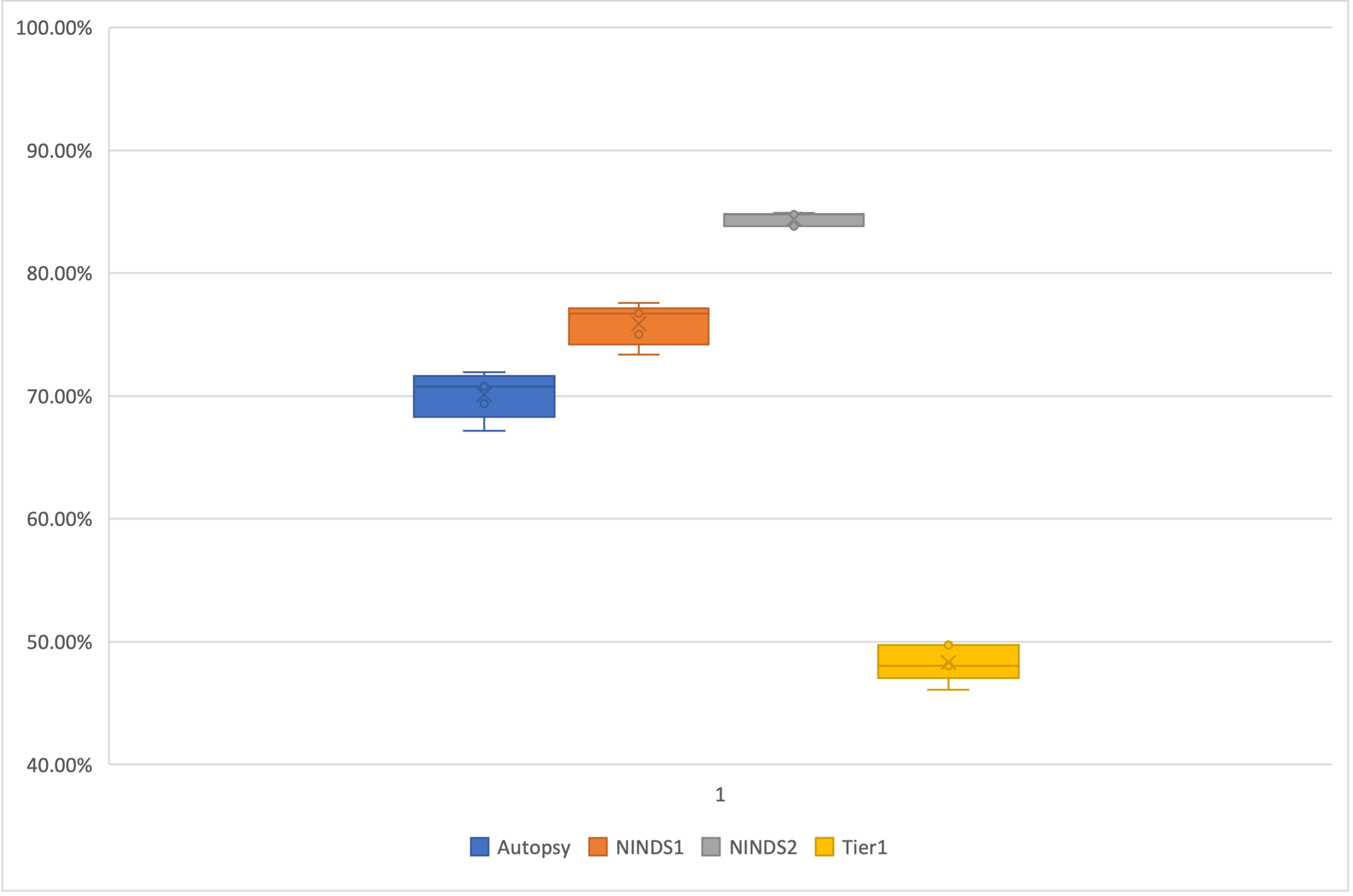}}
    \subfigure[Approach 1 (A)]{\includegraphics[width=0.45\textwidth]{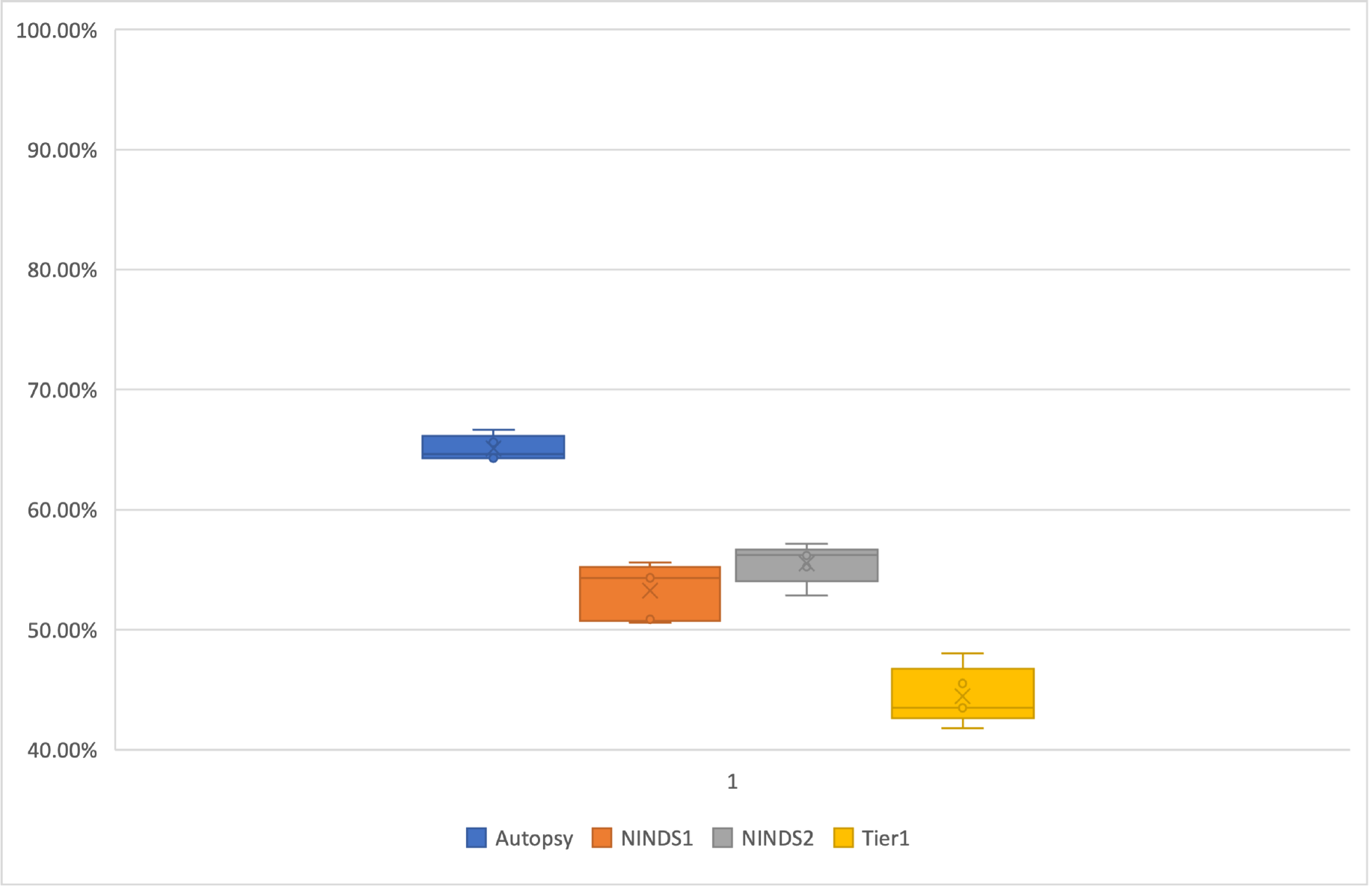}} 
    \subfigure[Approach 1 (B)]{\includegraphics[width=0.45\textwidth]{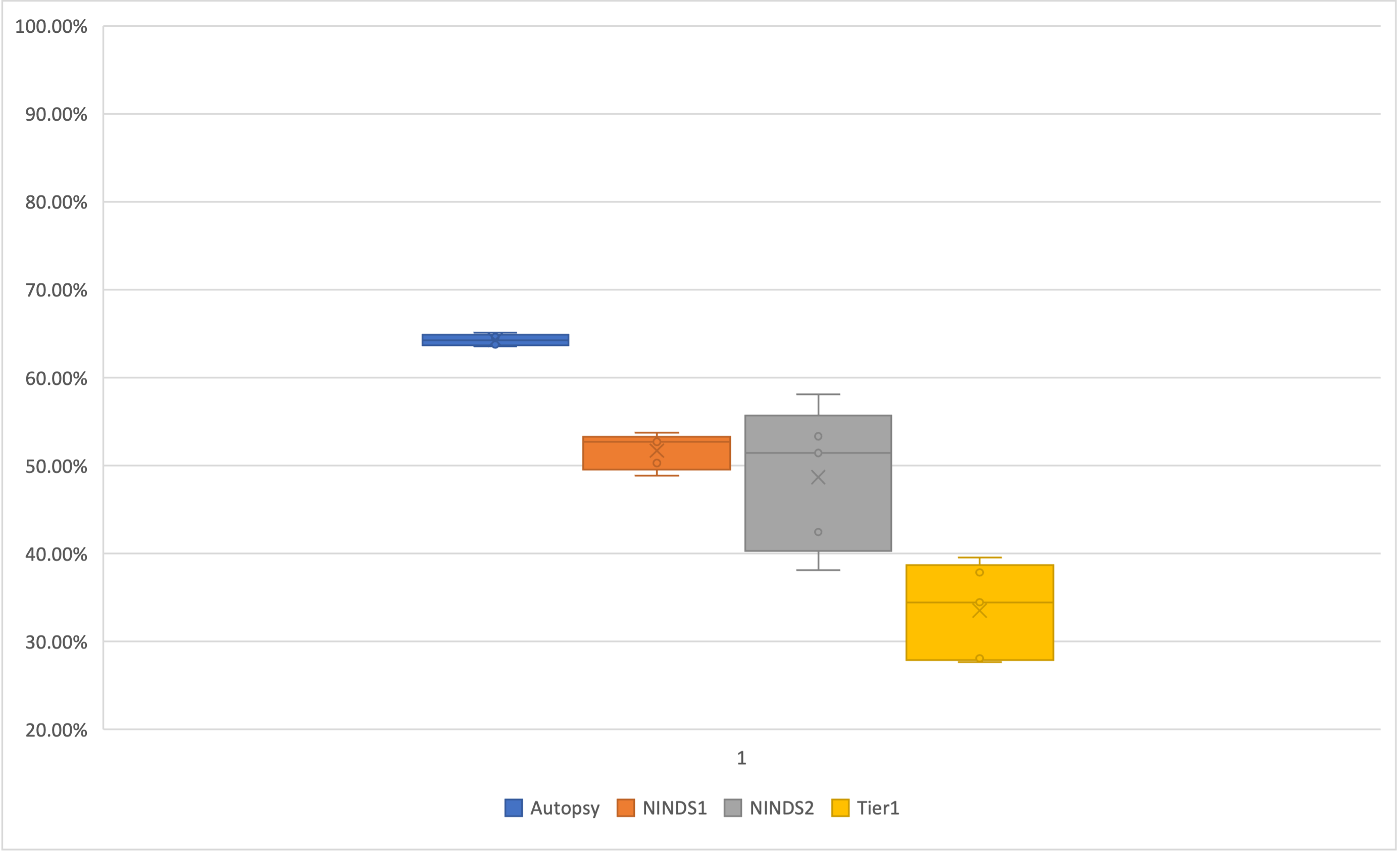}}
    \subfigure[Approach 2 (A)]{\includegraphics[width=0.45\textwidth]{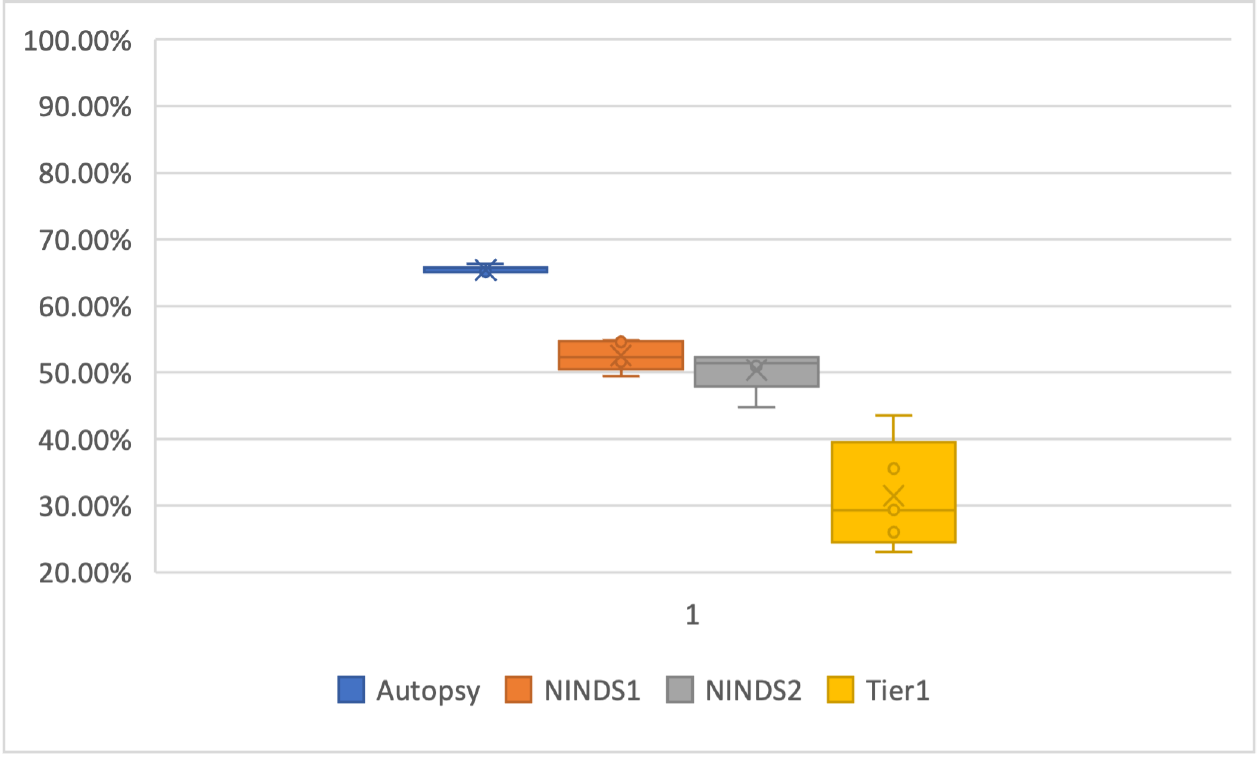}} 
    \subfigure[Approach 2 (B)]{\includegraphics[width=0.45\textwidth]{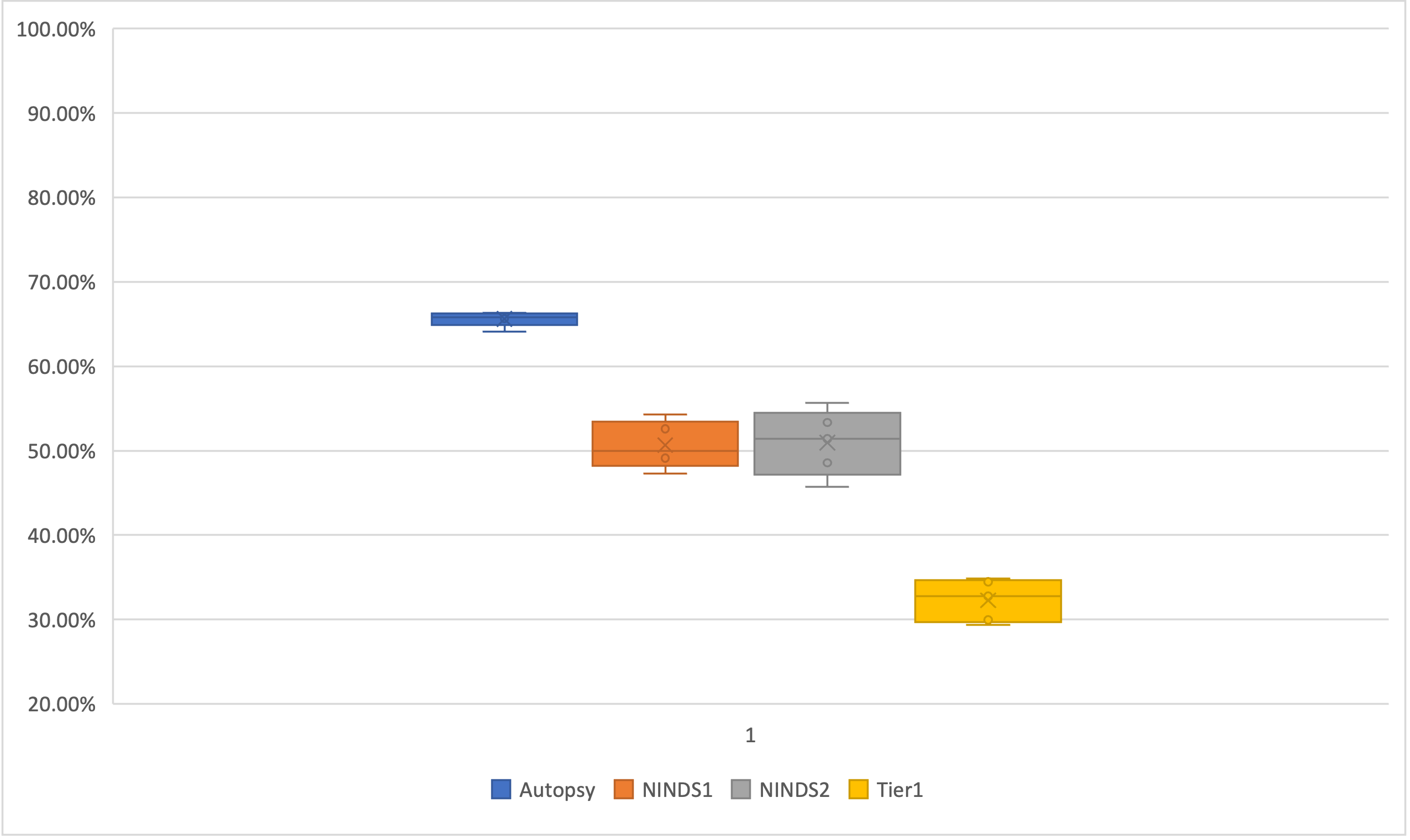}} 
    \caption{Accuracy for approaches 0, 1 and 2 per dataset}
    \label{All Approaches for datasets}
\end{figure}

\begin{figure}[h]
    \centering
    \subfigure[Approach 3 (A)]{\includegraphics[width=0.45\textwidth]{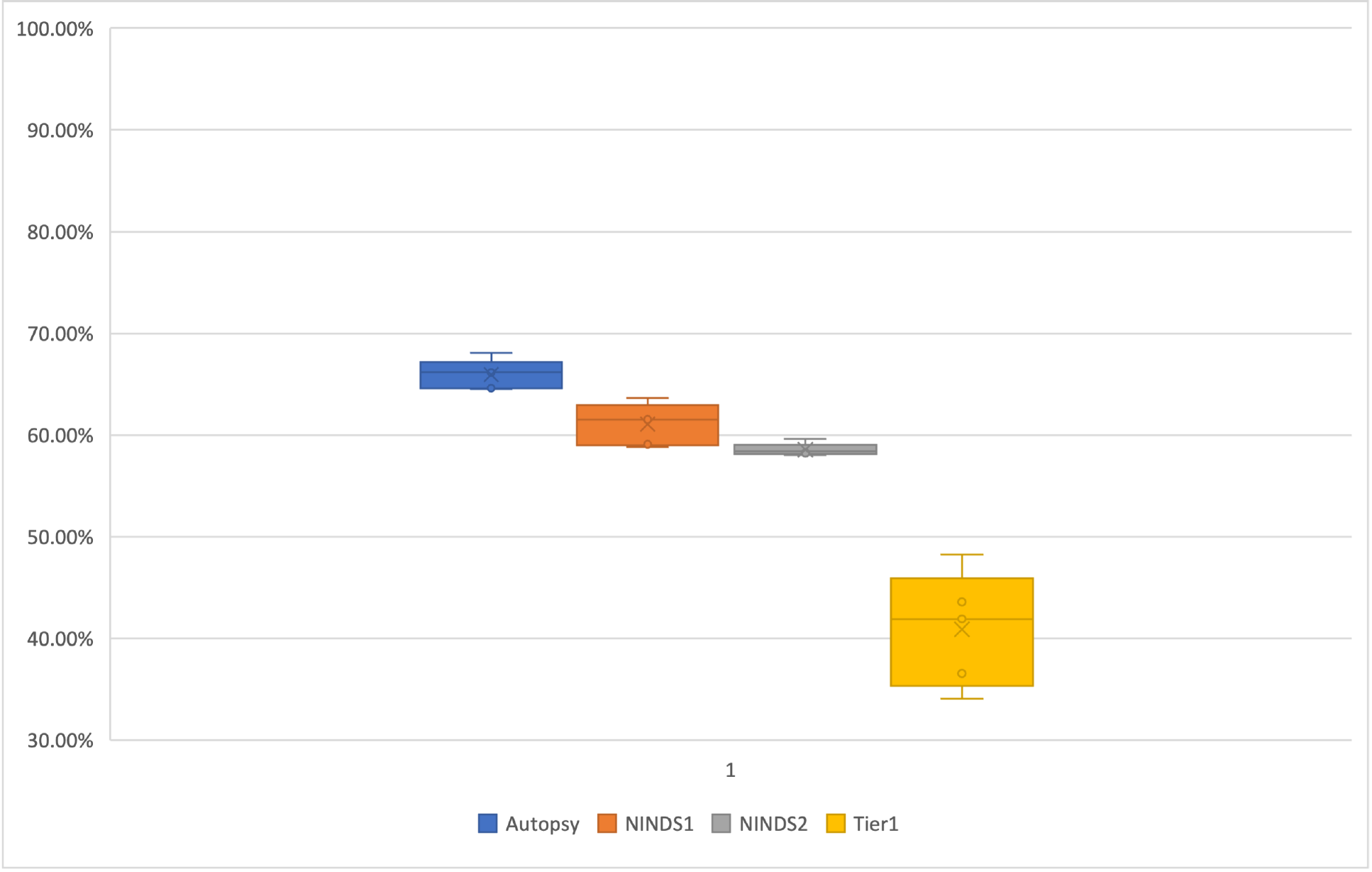}} 
    \subfigure[Approach 3 (B)]{\includegraphics[width=0.45\textwidth]{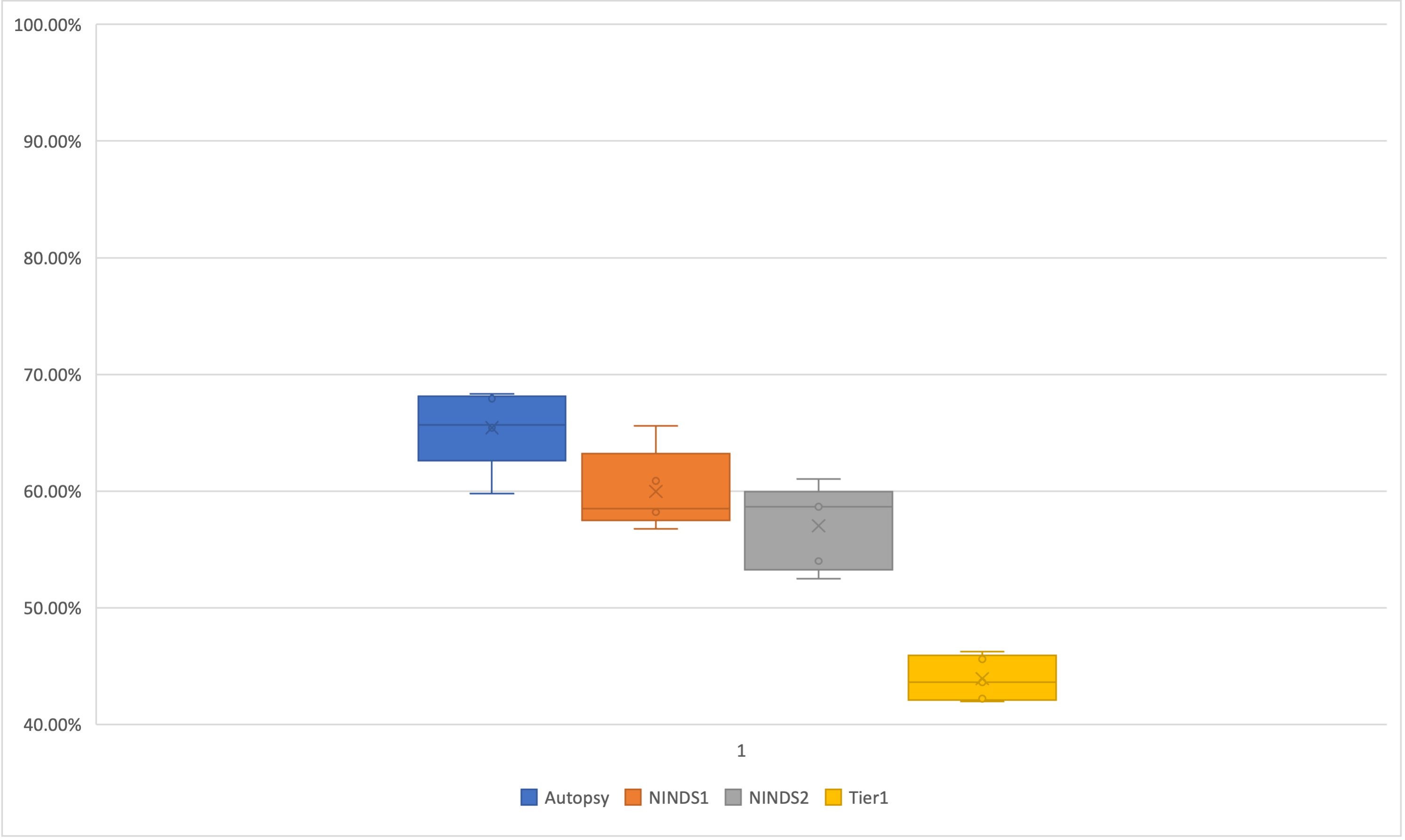}}
    \subfigure[Approach 4 (A)]{\includegraphics[width=0.45\textwidth]{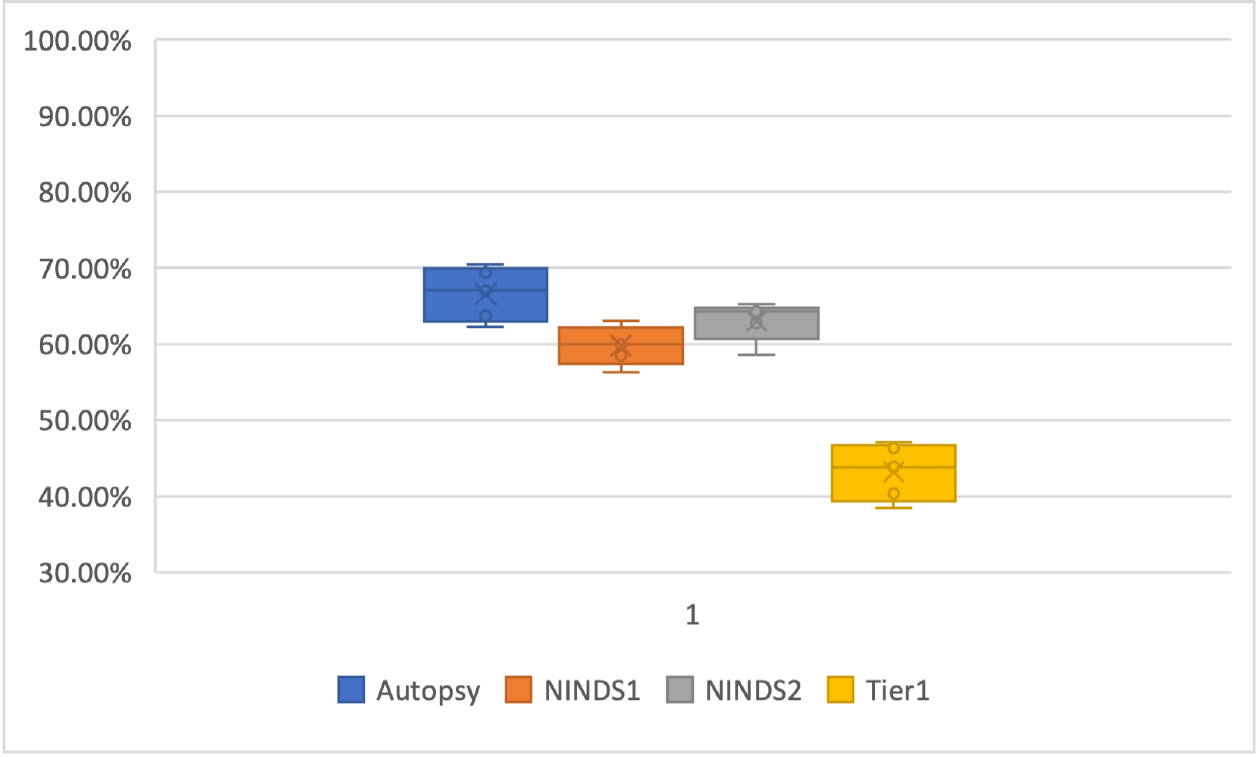}} 
    \subfigure[Approach 4 (B)]{\includegraphics[width=0.45\textwidth]{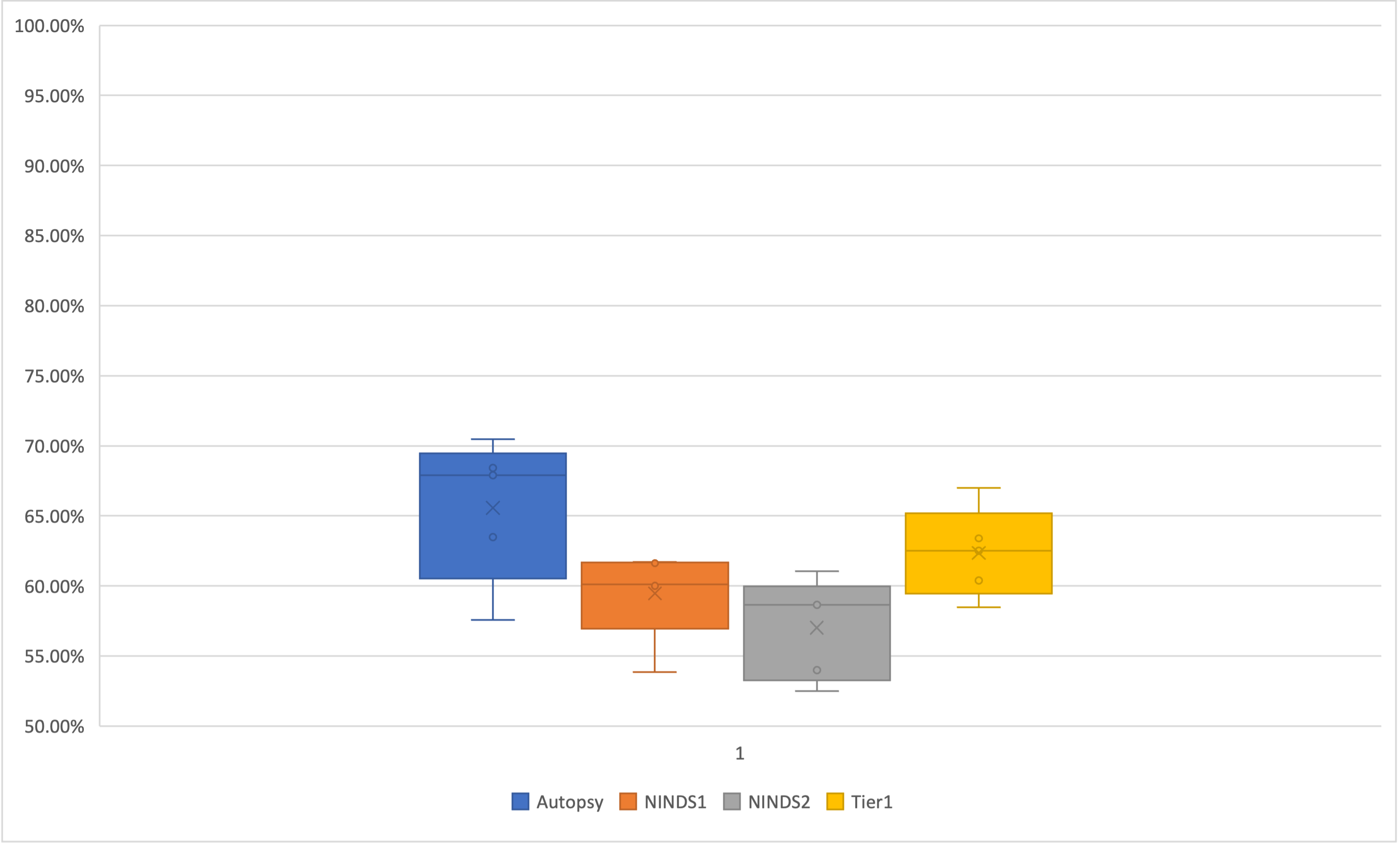}} 
    \caption{Accuracy for approaches 3 and 4 per dataset}
    \label{All Approaches for datasets 2}
\end{figure}

Approach 0 (A) obtained the highest accuracy for all datasets (Table~\ref{table:Comparison part} and Figures~\ref{All Approaches for datasets} and~\ref{All Approaches for datasets 2}). We anticipated that Approach 0 would achieve the highest accuracy because there is no technical variation among the samples used. Approaches 1 \& 2, and Approaches 3 \& 4 have similar accuracy.  Approaches 3 \& 4 achieved higher accuracy than approaches 1 \& 2  on Autopsy, NINDS1, and NINDS2. This lead us to recommend to perform feature selection on merged datasets. As extending the list of most common SNPs with SNPs in LD with the most common SNPs (mode B) did not improve accuracy, for the following analyses we considered only the SNPs obtained using mode A.

\begin{table}[h]
\centering
\caption{Comparison of datasets accuracy}
\label{table:Comparison part}
\resizebox{\columnwidth}{!}{%
\begin{tabular}{@{}lllll@{}}
\toprule
Approach & {$Autopsy \; Accuracy\pm sd$} & {$NINDS1 \; Accuracy\pm sd$} & {$NINDS2 \; Accuracy\pm sd$} & {$Tier1 \; Accuracy\pm sd$} \\
\midrule
Approach 0 (A) & $70.83\% \pm 1.74$ & $77.14\% \pm 2.12$ & $87.46\% \pm 3.30$ & $51.92\% \pm 3.12$ \\ 
Approach 1 (A) & $64.89\% \pm 0.20$ & $51.98\% \pm 2.87$ & $53.04\% \pm 4.83$ & $44.47\% \pm 2.13$ \\ 
Approach 2 (A) & $65.40\% \pm 0.47$ & $52.56\% \pm 2.02$ & $50.38\% \pm 2.86$ & $31.50\% \pm 7.31$ \\ 
Approach 3 (A) & $65.94\% \pm 1.30$ & $61.06\% \pm 1.85$ & $58.56\% \pm 0.55$ & $40.88\% \pm 5.05$ \\ 
Approach 4 (A) & $66.56\% \pm 3.16$ & $59.79\% \pm 2.31$ & $63.02\% \pm 2.36$ & $43.20\% \pm 3.36$ \\ 
Approach 0 (B) & $70.11\% \pm 1.69$ & $75.88\% \pm 1.52$ & $84.41\% \pm 0.49$ & $48.31\% \pm 1.35$ \\ 
Approach 1 (B) & $64.28\% \pm 0.56$ & $51.69\% \pm 1.83$ & $48.68\% \pm 7.33$ & $33.53\% \pm 4.89$ \\ 
Approach 2 (B) & $65.61\% \pm 0.79$ & $50.66\% \pm 2.50$ & $50.94\% \pm 3.50$ & $32.28\% \pm 2.25$ \\ 
Approach 3 (B) & $65.43\% \pm 3.06$ & $59.97\% \pm 3.10$ & $57.02\% \pm 3.22$ & $43.93\% \pm 1.74$ \\ 
Approach 4 (B) & $65.57\% \pm 4.60$ & $59.47\% \pm 2.89$ & $62.35\% \pm 2.88$ & $46.86\% \pm 3.90$ \\ 
 \bottomrule
\end{tabular}%
}
\end{table}

\begin{figure}[h]
    \centering
    \subfigure[Autopsy(SNPs)]{\includegraphics[width=0.24\textwidth]{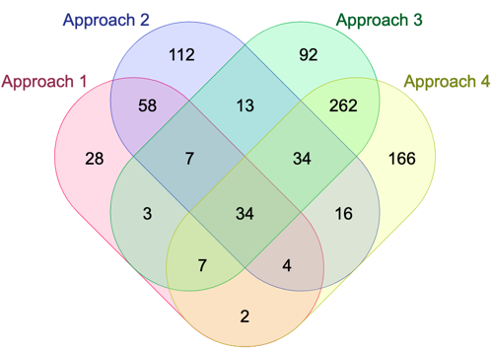}} 
    \subfigure[Autopsy(Genes)]{\includegraphics[width=0.24\textwidth]{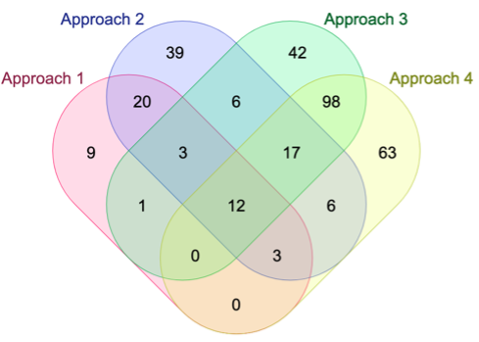}} 
    \subfigure[NINDS1(SNPs)]{\includegraphics[width=0.24\textwidth]{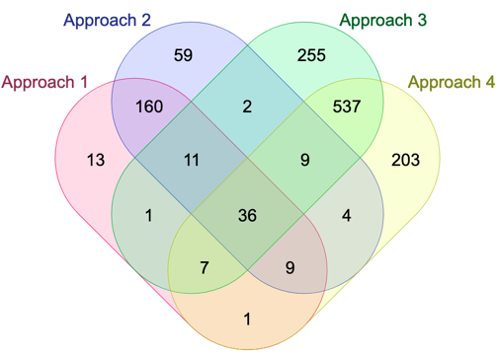}}
    \subfigure[NINDS1(Genes)]{\includegraphics[width=0.24\textwidth]{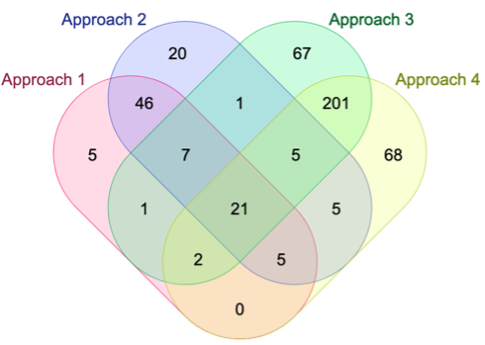}}
    \subfigure[NINDS2(SNPs)]{\includegraphics[width=0.24\textwidth]{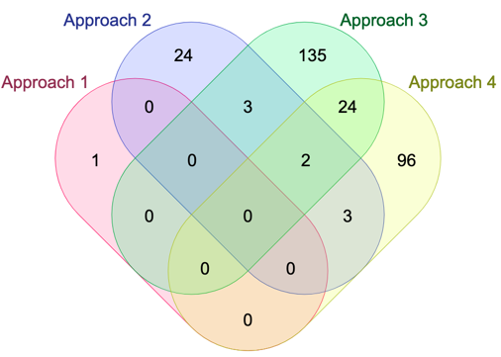}}
    \subfigure[NINDS2(Genes)]{\includegraphics[width=0.24\textwidth]{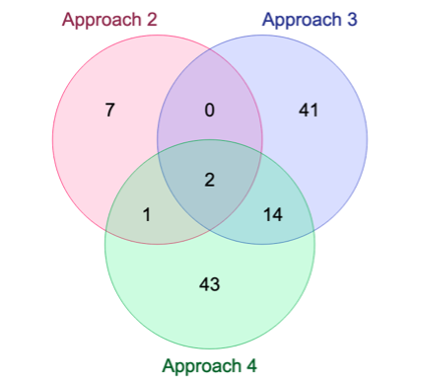}}
    \subfigure[Tier1(SNPs)]{\includegraphics[width=0.24\textwidth]{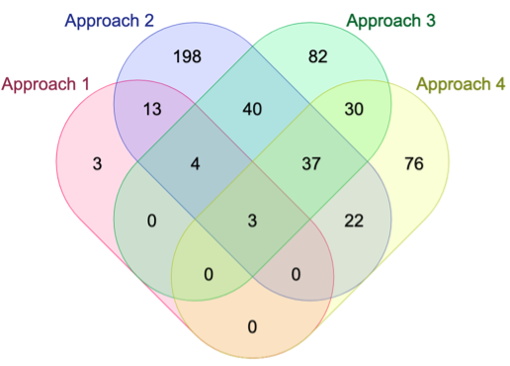}}
    \subfigure[Tier1(Genes)]{\includegraphics[width=0.24\textwidth]{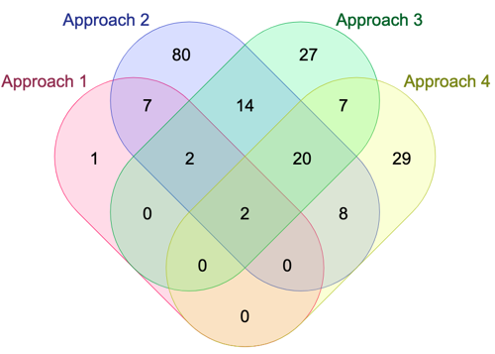}}
    \caption{Venn Diagram for Different approaches and same dataset. Venn diagrams were generated using molbiotools web site (https://molbiotools.com/listcompare.php).}
    \label{fig:Venn1}
\end{figure}

In Figure~\ref{fig:Venn1}, we show the number of SNPs and genes that are in common among different integration approaches for the same dataset. To get the genes where the SNPs are located we used the Biomart platform \citep{biomart}.   On average around 6\%  of the SNPs identified using Approach 0 (without data  integration) are  replicated by at least one other approach (Supplementary Figures), while on average $38.1 \pm 22$\% of the SNPs identified with any data integration strategy (Approaches 1 to 4) are replicated by at least one other integration approach (Figure~\ref{fig:Venn1}). This indicates that integrating datasets  increases by six-fold the proportion of replicated SNPs.  The highest agreement was observed between Approaches 3 and 4, followed by Approaches 1 and 2. Figure~\ref{fig:Venn1}(f) only includes three approaches because there were no common SNPs between Approach 1 and Approaches 2, 3, and 4.

\begin{figure}[h]
    \centering
    \subfigure[Approach 0(SNPs)]{\includegraphics[width=0.24\textwidth]{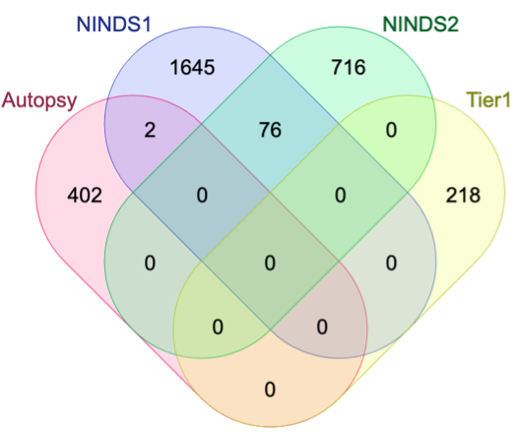}} 
    \subfigure[Approach 0(Genes)]{\includegraphics[width=0.24\textwidth]{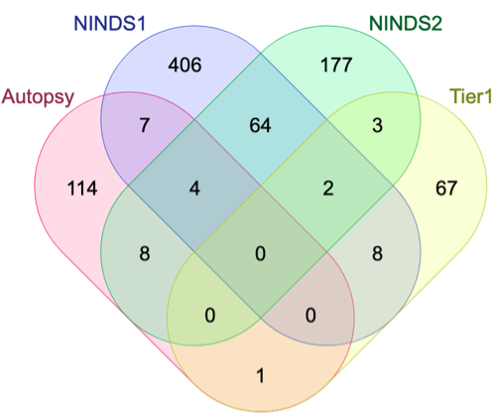}} 
    \subfigure[Approach 1(SNPs)]{\includegraphics[width=0.24\textwidth]{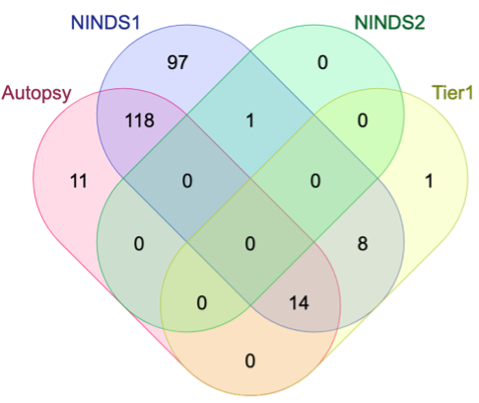}}
    \subfigure[Approach 1(Genes)]{\includegraphics[width=0.24\textwidth]{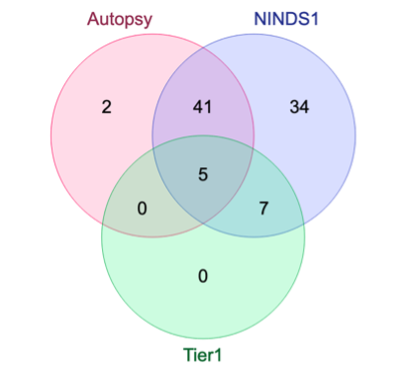}}
    \subfigure[Approach 2(SNPs)]{\includegraphics[width=0.24\textwidth]{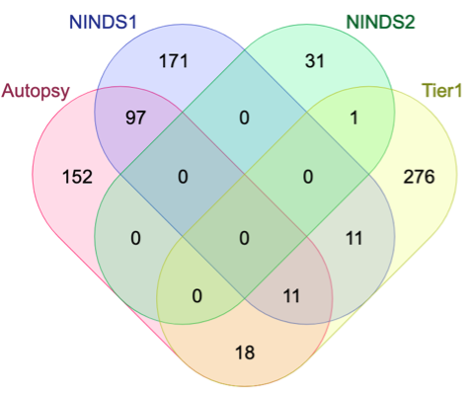}}
    \subfigure[Approach 2(Genes)]{\includegraphics[width=0.24\textwidth]{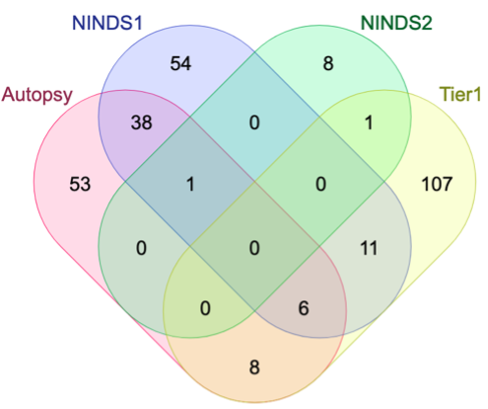}}
    \subfigure[Approach 3(SNPs)]{\includegraphics[width=0.24\textwidth]{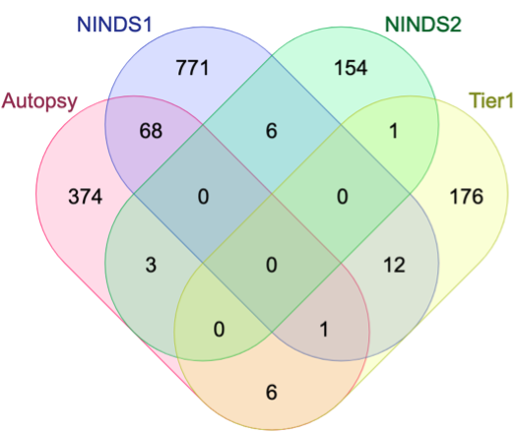}}
    \subfigure[Approach 3(Genes)]{\includegraphics[width=0.24\textwidth]{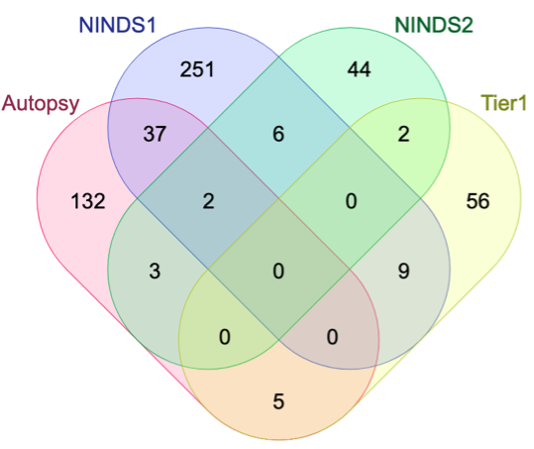}}
    \subfigure[Approach 4(SNPs)]{\includegraphics[width=0.24\textwidth]{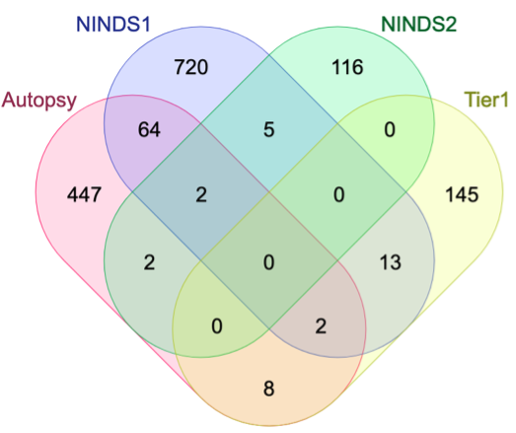}}
    \subfigure[Approach 4(Genes)]{\includegraphics[width=0.24\textwidth]{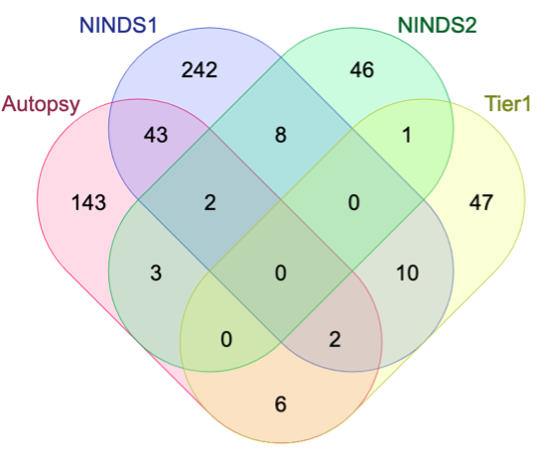}}
    \caption{Venn Diagram for Different datasets and same approach. Venn diagrams were generated using molbiotools web site (https://molbiotools.com/listcompare.php).}
    \label{fig:Venn2}
\end{figure}

Figure~\ref{fig:Venn2} shows the number of identified SNPs and genes that are in agreement using different datasets per approach. The number of SNPs in agreement between datasets fell dramatically when using different datasets but the same approach. The highest pair-wise agreement is in Approach 0 between NINDS1 and NINDS2 which are two datasets generated by the same studies. This suggests that the genotyping platform and study design are the greatest source of variation. If we consider NINDS1 and NINDS2 as the same dataset then the number of genes reproduced by at least two different datasets goes from 33 in Approach 0 to 75 in Approach 4. This suggests that integrating datasets more than duplicates the number of genes that are reproduced in at least two datasets.

\subsection{Phenotypes associated with potential biomarkers}\label{biomarkers and associated phenotypes}
To investigate whether SNPs identified by at least two datasets or two approaches (referred to as replicated SNPs) are associated with PD, we retrieved the phenotype associated with each of these replicated SNPs using the Biomart platform.  Four of these replicated SNPs (rs11248060, rs239748, rs999473, and rs2313982) are directly linked to PD. By performing a literature search on the associated phenotypes, we found 50 other replicated SNPs indirectly associated with PD. That is, they are directly associated with a disease that frequently co-occurs with PD.  These 50 replicated SNPs together with the manuscripts used as basis of their indirect association to PD are listed in Supplementary File 1. These 50 replicated SNPs open up new potential avenues of investigation for PD biomarkers.

\section{Conclusion}\label{sec6}

The integration of datasets may result in a slight decrease in classification accuracy but it can significantly enhance the reproducibility of potential biomarkers.    We show that on average 93\% of the SNPs discovered using a single data set are not replicated in other datasets, and  that dataset integration reduces that average lack of replication to 62\%. A limitation of this study is that SNPs genotyped on each dataset differ due to different genotyping platforms, leading to the exclusion of many SNPs as potential biomarkers when integrating data sets. To address this issue, repeating the study on SNPs detected by whole-genome sequencing could be a solution. In addition, accessing other datasets from diverse populations would enable us to further test the replication of our identified SNPs. Moreover, our methodology is applicable to various diseases such as breast cancer, lung cancer, and colorectal cancer.

\section*{Declaration of  Interest}
None.

\section*{Funding Sources}

This research was partially supported by grants from the Natural Sciences and Engineering Research
Council of Canada (NSERC) to L.P.-C. and H.U.  A.A was partially supported by funding from Memorial University School of Graduate Studies.

\section*{Declaration of generative AI in scientific writing}
During the preparation of this work the author(s) used Grammarly and ChatGPT in order to check the grammar and improve clarity.  After using this tool/service, the author(s) reviewed and edited the content as needed and take(s) full responsibility for the content of the publication.

\bibliographystyle{elsarticle-num} 

\end{document}